%% file: ICRC2023_CamSim.tex
\documentclass[a4paper,11pt]{article}

\usepackage{pos}
\usepackage{lipsum} 
\usepackage{caption}
\usepackage{subcaption}

\title{A new simulation framework for IceCube Upgrade calibration using IceCube Upgrade Camera system}

\ShortTitle{CamSim}

\author{The IceCube Collaboration \\{\normalsize \normalfont(a complete list of authors can be found at the end of the proceedings)}\\}

\emailAdd{rott@physics.utah.edu}

\abstract{
Currently, an upgrade consisting of seven densely instrumented strings in the center of the volume of the IceCube detector with new digital optical modules (DOMs) is being built. On each string, DOMs will be regularly spaced with a vertical separation of 3 m between depths of 2160 m and 2430 m below the surface of the ice, which is a denser configuration than the existing DOMs of IceCube detector.

For a precise calibration of the IceCube Upgrade it is important to understand the properties of the ice, both inside and surrounding the deployment holes. The camera system together with the LED illumination system was developed and produced at Sungkyunkwan university and are installed in almost every DOM to measure these properties. For these calibration measurements, a new simulation framework, which produces expected images from various geometric and optical variables has been developed.  Images produced from the simulation will be used to develop an analysis framework for the IceCube Upgrade camera calibration system and for the design of the IceCube Gen2 camera system.

\vspace{4mm}
{\bfseries Corresponding authors:}
Carsten Rott$^{1,2}$, Christoph T\"onnis$^{1}$, Seowon Choi$^{1,*}$, Jiwoong Lee$^{1}$, Minyeong Seo$^{1}$\\
{$^{1}$ \itshape Department of Physics, Sungkyunkwan University, Rep. of Korea}\\
{$^{2}$ \itshape Department of Physics and Astronomy, University of Utah, Salt Lake City, UT 84112, USA}\\
\\[4mm]
$^*$ Presenter

\ConferenceLogo{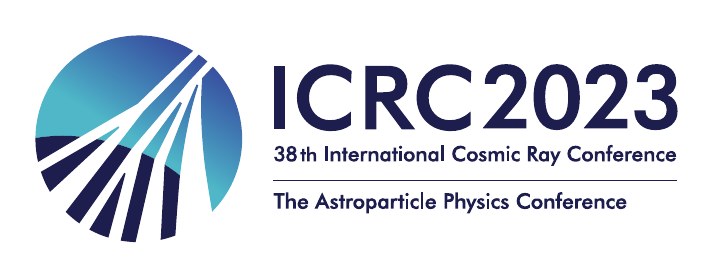}

\FullConference{The 38th International Cosmic Ray Conference (ICRC2023)\\ 26 July -- 3 August, 2023\\ Nagoya, Japan}
}

\begin{document}

\maketitle

\section{Introduction}\label{sec1}

The IceCube Upgrade~\cite{ICRC2019:ICU-project} is an extension to the IceCube detector~\cite{ICdetector} that adds 7 new strings with novel digital optical modules (DOMs)~\cite{ICRC2021:D-Egg,ICRC2021:mDOM} in the center of the detector volume to increase the sensitivity of IceCube to low energy neutrinos and to use new devices to calibrate the IceCube detector. A key part of this calibration effort is the measurement of properties of the Antarctic Ice in the detector volume, such as quantifying an anisotropy of the optical properties of the ice~\cite{ICRC2019:anisotropy}, and refine existing models for the Ice~\cite{SPICE}. To this end a new camera system has been developed for integration into the new DOMs. 

A general simulation framework based on the Photon Propagation Code (PPC) has been developed for these cameras. This framework is capable of simulating generic optical properties of homogeneous or layered media, generic light sources with gaussian profiles and cameras with arbitrary resolutions, fields of view and lenses. The new framework is highly versatile and is not only being used to develop analyses for the expected data of the camera system in the IceCube Upgrade~\cite{ICRC2019:ICUcamera} but also for design studies for a system for IceCube Gen2~\cite{ICRC2017:Gen2camera} and other future systems for use in particle detectors in ice and water.

\section{The CamSim framework}\label{sec2}

\begin{figure}[h!]
    \centering
   \includegraphics[width=.9\linewidth]{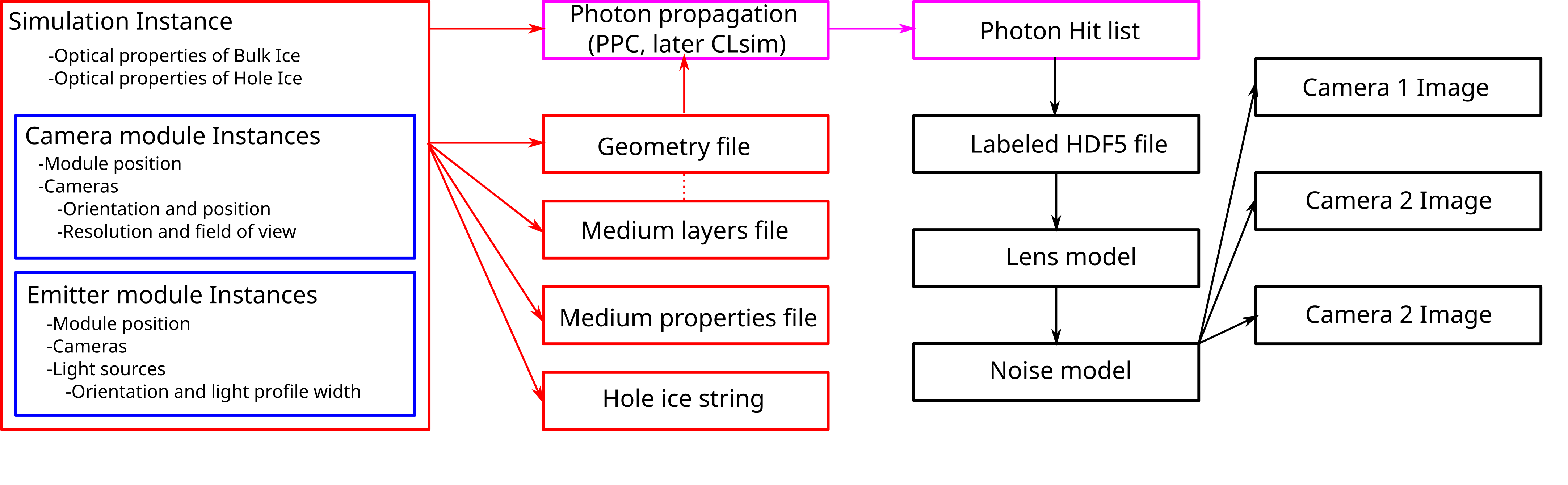}
    \caption[margin=0.8cm]{An overview of the CamSim Framework. Parts in Blue and Red are part of the CamSim wrapper, Purple indicates the photon propagation code, Black are output processing by the CamSim wrapper.}
    \label{fig:CamSim}
\end{figure}

The core of the simulation framework is a program called PPC~\cite{ppc}. PPC is a C and Cuda based code that was designed to simulate photon propagation inside the IceCube detector. The program is set up to simulate the entire detector and is inflexible for use with cameras. For CamSim PPC was modified to be able to simulate the different light sources used and a python-based wrapper script was created to handle the complexities of PPC and provide a simplified interface for the simulation. It also improves on PPC by organizing output data and including functions to convert the output into images.The framework is planned to include other simulation codes to process photon propagation such as CLsim~\cite{Clsim:2022yy}. PPC was chosen to be included first due to the speed with which photon statistics can be generated. CLsim would allow to generate figures that show the path individual photons are taking in the simulation.

The PPC simulation code requires geometry and ice layer files to provide information on the position of the modules of the IceCube detector and the current knowledge layering of the medium in the detector. Further files for the simulation setup contain the angular acceptance of the PMTs in IceCube. For the camera simulations the geometry file is simplified to contain only a minimal number of modules needed for the simulation case, which in most cases is one module with a light source and one with one or more cameras. The ice layering is currently being neglected in the simulations as the camera measurements are expected to see photons from no more than a few 10 meters due to scattering and absorption. 

PPC has also the capacity to simulate a column of material of different optical properties. This feature aims to simulate the differing optical properties in the re-frozen ice in the deployment holes of IceCube. CamSim generates a string to parse this information to PPC based on parameters specified in the simulation instance. 

The flow of the CamSim framework is shown in Figure \ref{fig:CamSim}. Within a simulation instance of CamSim Emitter and Camera instances are created specifying the positions of modules with a 33 cm diameter corresponding to the size of the digital optical modules in IceCube Gen1 (called pDOMs). The Emitter instances each have one light source attached. Both types of instances can have cameras attached on any point on the module source. The simulation instance contains the parameters for the ice properties.

Cameras are simulated using a simple pinhole camera simulation by default, though additional models for lens effects are being implemented. The cameras are characterised by a horizontal and vertical resolution and a horizontal and vertical Field of View (FOV). Due to the nature of a simple pinhole camera we are restricting the FOV values in the simulation to 90 degrees. The resolution can be set to any arbitrary value. For the camera for the IceCube Upgrade the actual resolution is 1312 pixels horizontally by 979 pixels vertically. The image sensor has 3 types of pixels sensitive to red, green and blue light respectively. This image sensor has its pixels in 2 by 2 groups with one red, 2 green and one blue pixel in each group to capture images in color. The LEDs used for the upgrade are predominantly blue and in most simulations we are using a resolution of 500 by 500 pixels to reflect the approximately 500 by 500 blue pixels in the actual image sensor. 

\begin{figure}[h!]
    \centering
   \includegraphics[width=.7\linewidth]{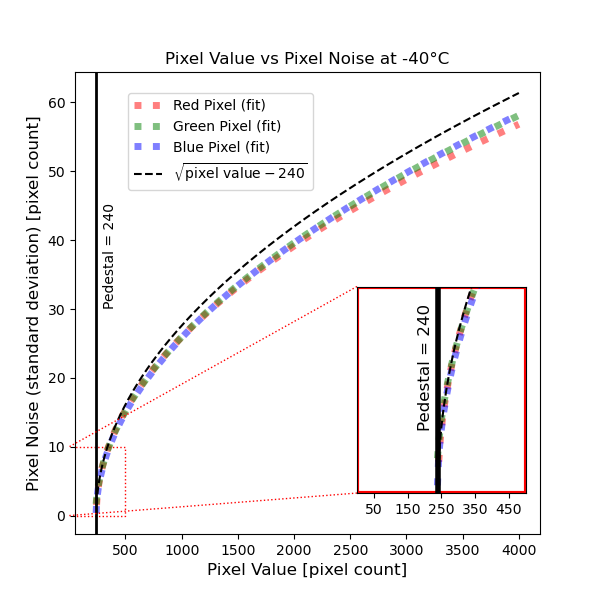}
    \caption[margin=0.8cm]{The noise model used for the camera system for the IceCube Upgrade. The noise behavior of the cameras follows a power law. This model was obtained from measurements at Sungkyunkwan University by evaluating a large number of images captured under stable illumination at -40 degrees Celsius with an exposure time of 3.7 seconds. In the analysis different color types of pixels were treated separately, but showed near identical behavior.}
    \label{fig:camera_noise}
\end{figure}

These simulated images are used within an analysis framework to develop sensitivities and likelihood functions for the camera system. In the analysis framework a noise model for the camera, shown in Figure \ref{fig:camera_noise}, is applied. The noise model was produced on long-time measurements of 20 cameras at -40 degrees Celsius. The cameras were illuminated with a reference light source through a light diffuser. The variation of the pixel response was then evaluated as a function of the average brightness of the pixel. The response of a pixel to illumination is called pixel count, which is the integer digital value that is read out from the image sensor. The variation $s$ is related to the average brightness $n$ approximately by $s = \sqrt{n-240}$, where 240 is the readout a pixel gives when it is not illuminated.

\begin{figure}[h!]
    \centering
   \includegraphics[width=.49\linewidth]{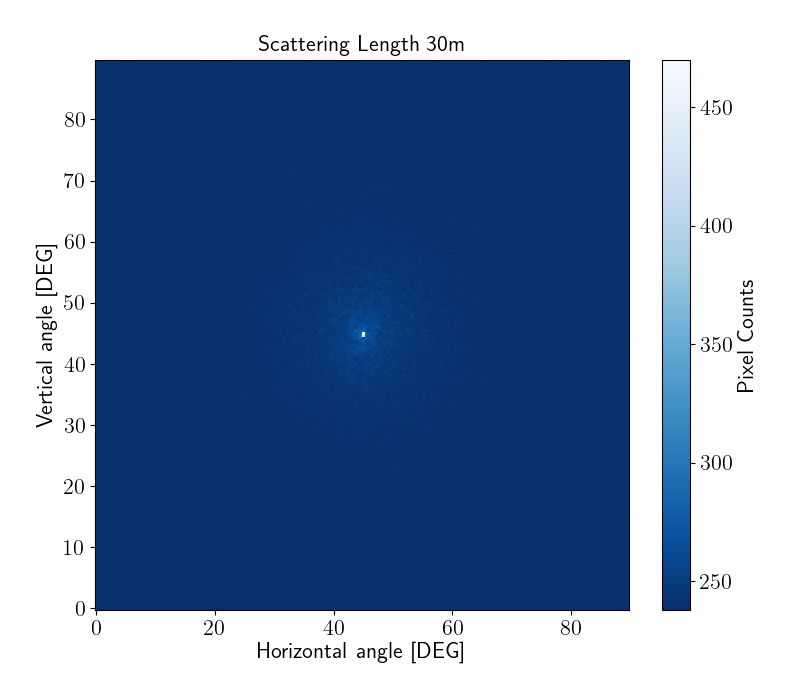}
    \includegraphics[width=.49\linewidth]{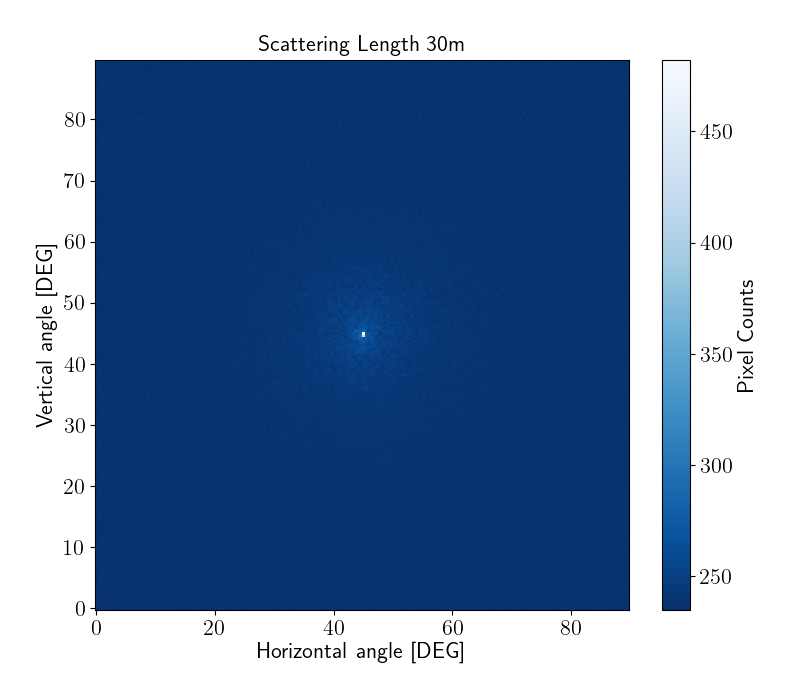}
    \caption[margin=0.8cm]{A sample image without noise (left) and with noise (right). The noise makes the image appear more granulated, but the overall scattering pattern is still well observable.}
    \label{fig:noiseimage}
\end{figure}

In Figure \ref{fig:noiseimage} a simulated image without noise (on the left) and with noise (on the right) is shown. The image with noise is slightly more grainy, though the effect is not easily apparent with bare eye. 

\section{Hole Ice simulations}\label{sec3}

The simulation studies on the refrozen hole ice for the IceCube Upgrade are focused on determining the size, position and the scattering and absorption lengths of a potential column of bubbles that had first been detected by a special camera system deployed below the deepest DOM of IceCube string~80~\cite{ICdetector}. These bubble columns and the general ice properties in the drillholes are a major source of systematic uncertainty for many IceCube analyses such as oscillation studies~\cite{Aartsenosc}.

\begin{figure}[ht]
    \centering
   \includegraphics[width=.7\linewidth]{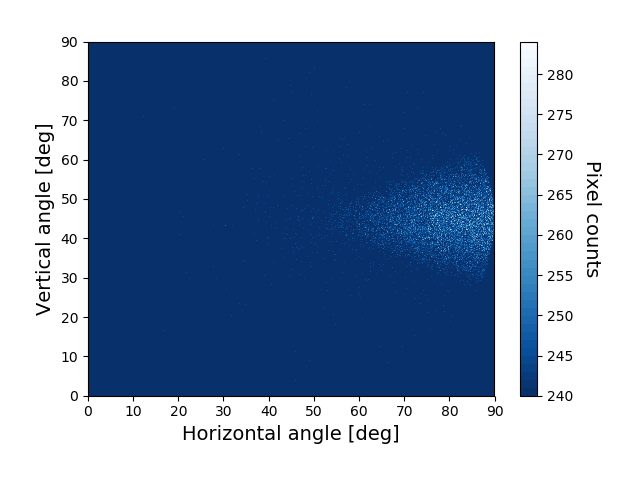}
    \includegraphics[width=.29\linewidth]{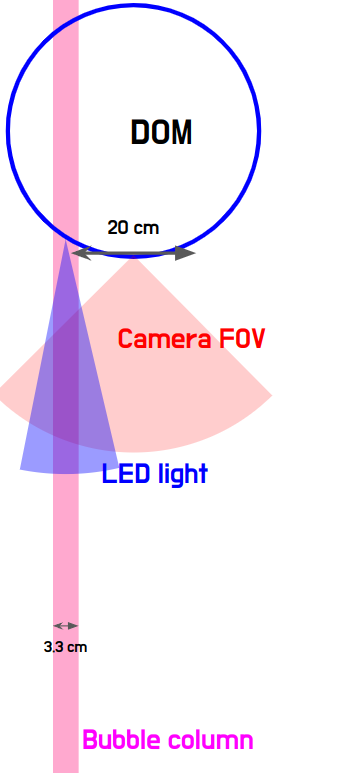}
    \caption[margin=0.8cm]{A sample simulated image of the hole ice seen looking down. The simulated bubble column is visible in the image to the left of the camera. The diameter of the column here is 3.3 cm.}
    \label{fig:hole_ice}
\end{figure}

Existing calibration efforts have been able to limit the size and scattering length of the bubble column using measurements with the LED flasher system in IceCube~\cite{refId0}, however these measurements mostly estimate the total amount of light scattered or absorbed in the column. The new camera system is expected to be able to measure these parameters independently and precisely and complement the existing measurements.

An earlier simulation of this bubble column is seen in Figure \ref{fig:hole_ice}. This simulation was using a camera field of view of 90 degrees and is looking down from a simulated DOM with the bubble column offset from the camera’s optical axis by 10 cm. The scattering length in this bubble column is 5 cm. A sketch of the geometry is shown on the right side of the figure. 

Based on the properties of the simulated bubble column the position, shape and brightness of the visible light in the simulated image changes. 

\section{Geometry and bulk ice simulations}\label{sec4}

The optical properties of the ice between the IceCube strings is described through its absorption and scattering length. The main purpose of the bulk ice simulation studies is to create a framework to measure the scattering length and absorption length using the camera system deployed into the ice. For these measurements we will use LED light captured by cameras in adjacent strings to the emitter. As the strings of the IceCube Upgrade are placed more densely than the existing IceCube strings, it is expected that the scattering and absorption length can be measured with greater accuracy.

\begin{figure}[ht]
    \centering
    \begin{subfigure}{\linewidth}
        \centering
        \includegraphics[width=.48\linewidth]{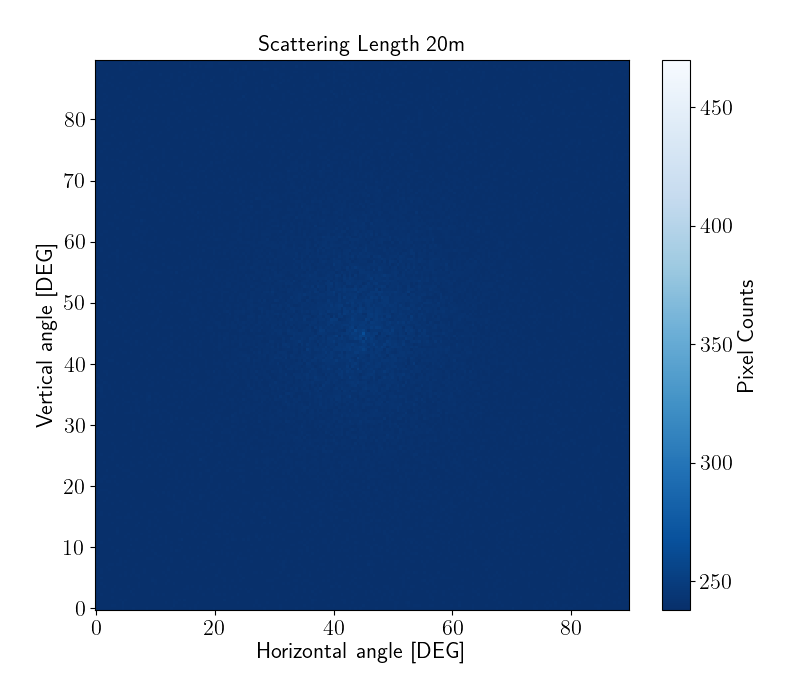}
        \includegraphics[width=.48\linewidth]{figures/sc30_np12_scaled.png}
        \caption[margin=0.8cm]{Simulated images with scattering length 20 m and 30 m}
    \end{subfigure}
    \begin{subfigure}{.70\linewidth}
        \includegraphics[width=.75\linewidth]{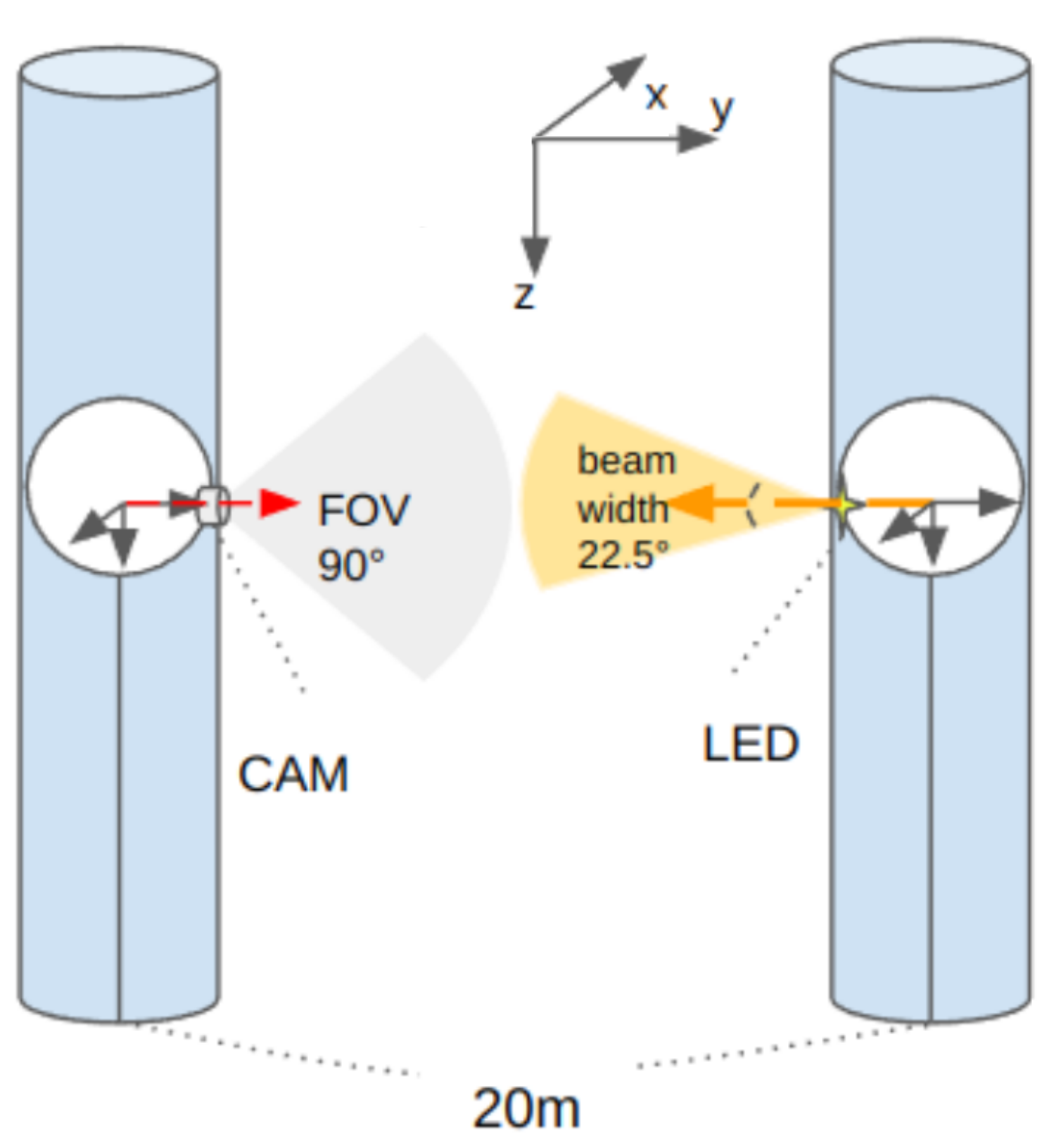}
        \caption[margin=0.8cm]{Schematics of simulation (a). Not to scale.}
    \end{subfigure}
    \caption[margin=0.8cm]{{\bf(a)} Simulated images with varying scattering length {\bf(b)} Geometry of the bulk ice measurement.}
    \label{fig:enter-label}
\end{figure}

Current simulation efforts have been generating images for different scattering lengths and different orientations of the LED with regards to the camera. As can be seen in Figure \ref{fig:enter-label} the visible light cone from the LED is very apparently different for variations in scattering length from 20 m to 30 m.

\begin{figure}
        \centering
        \includegraphics[width=.45\linewidth]{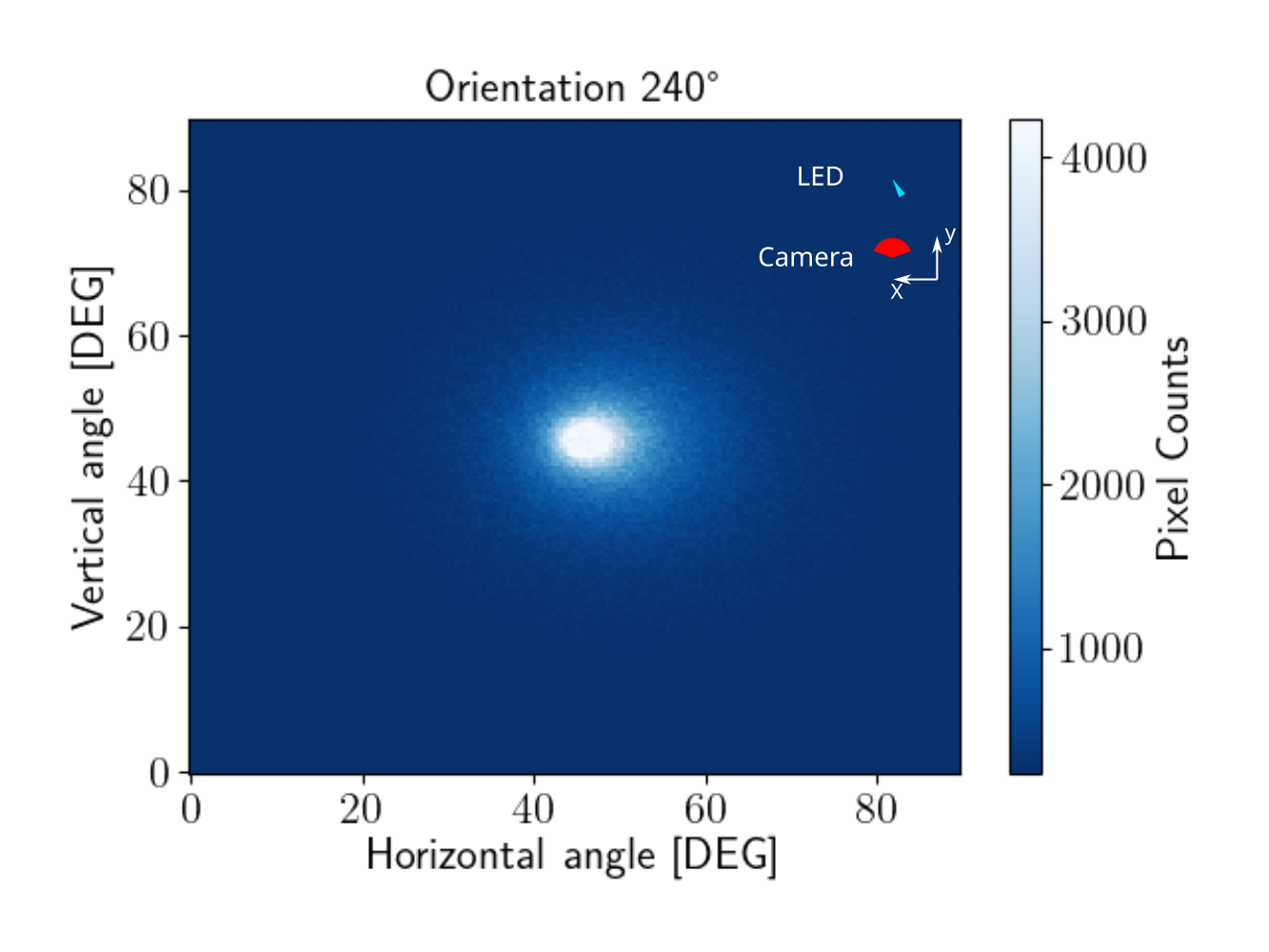}
        \includegraphics[width=.45\linewidth]{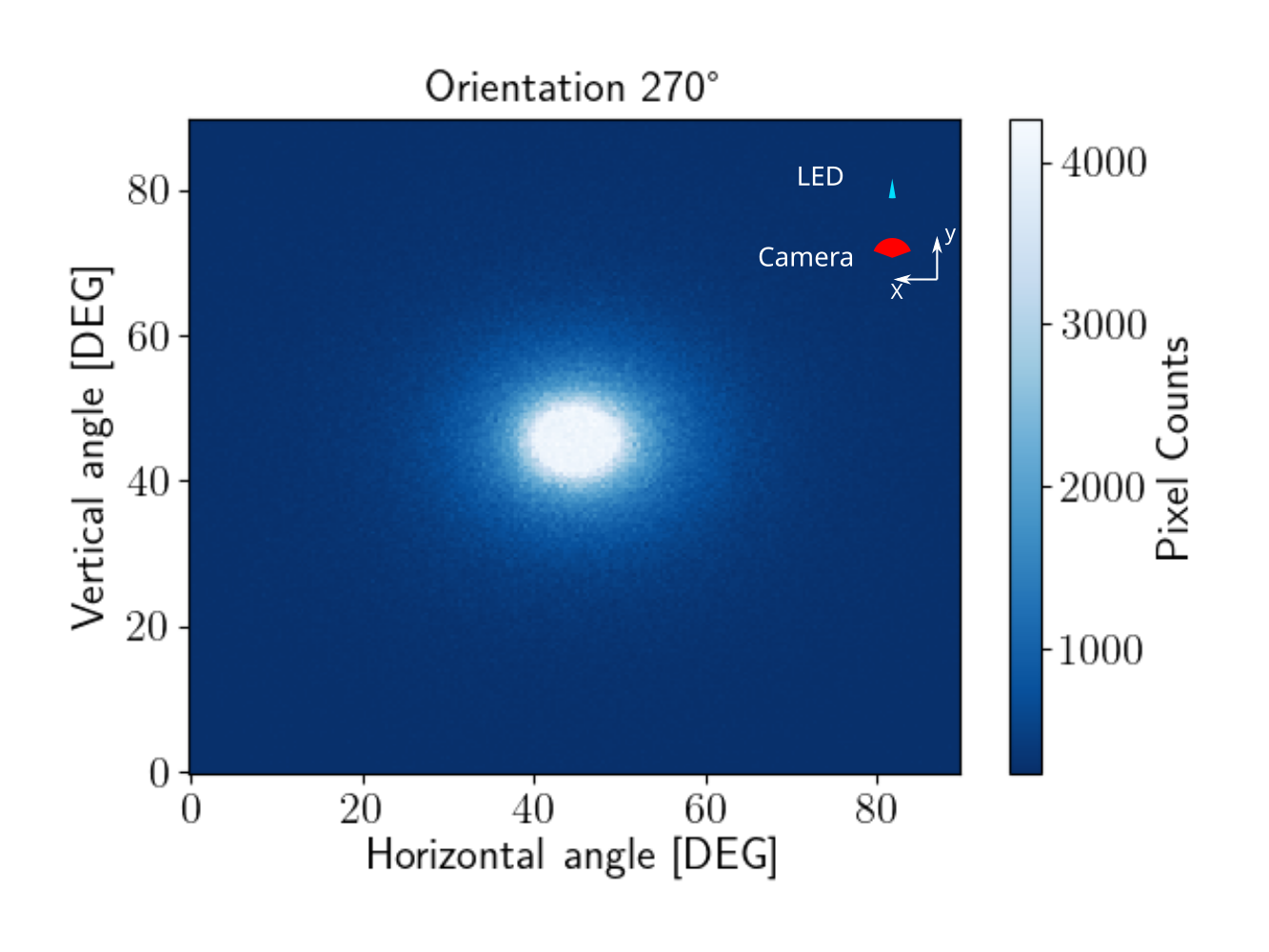}
        \includegraphics[width=.45\linewidth]{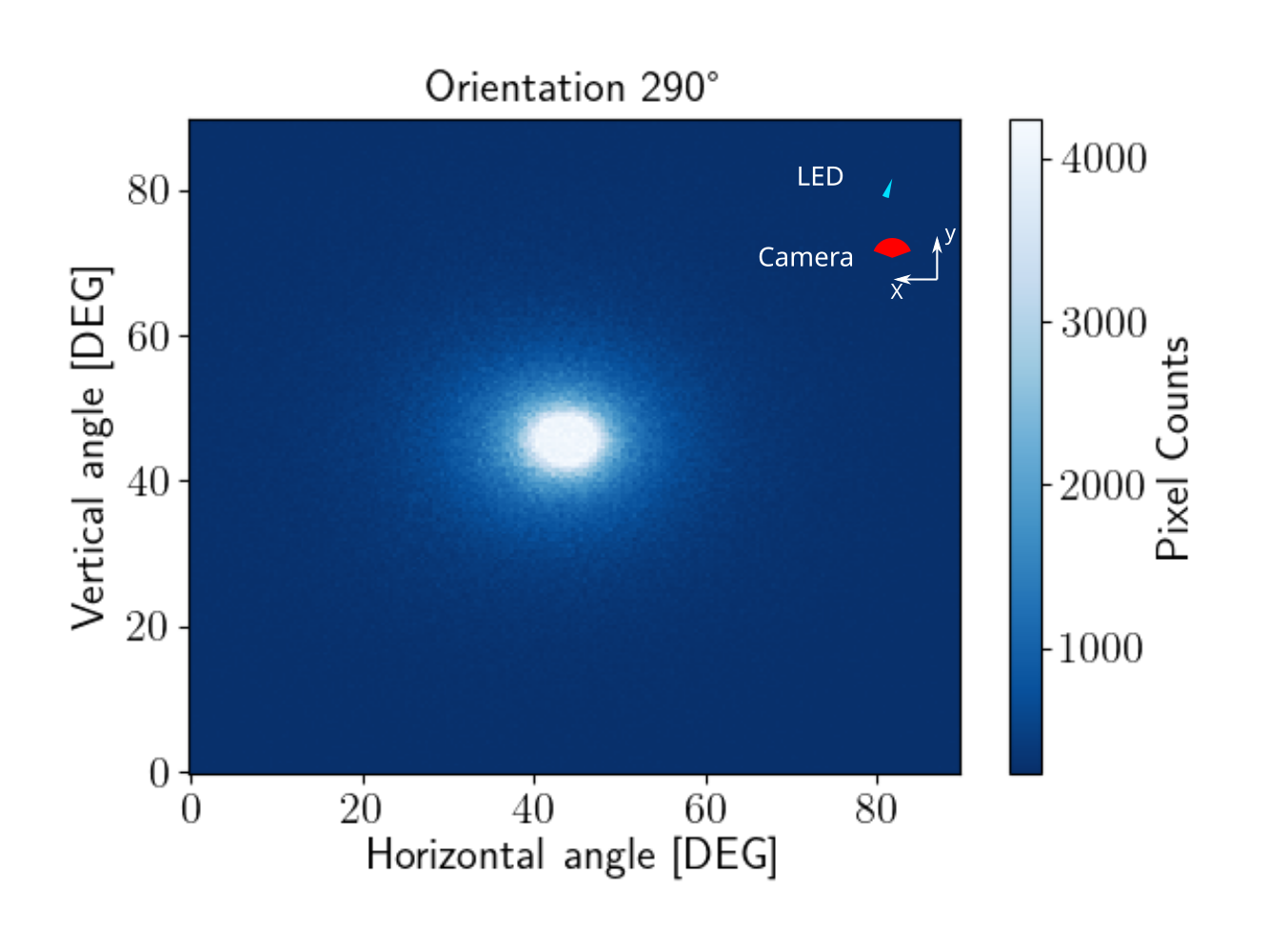}
        \includegraphics[width=.45\linewidth]{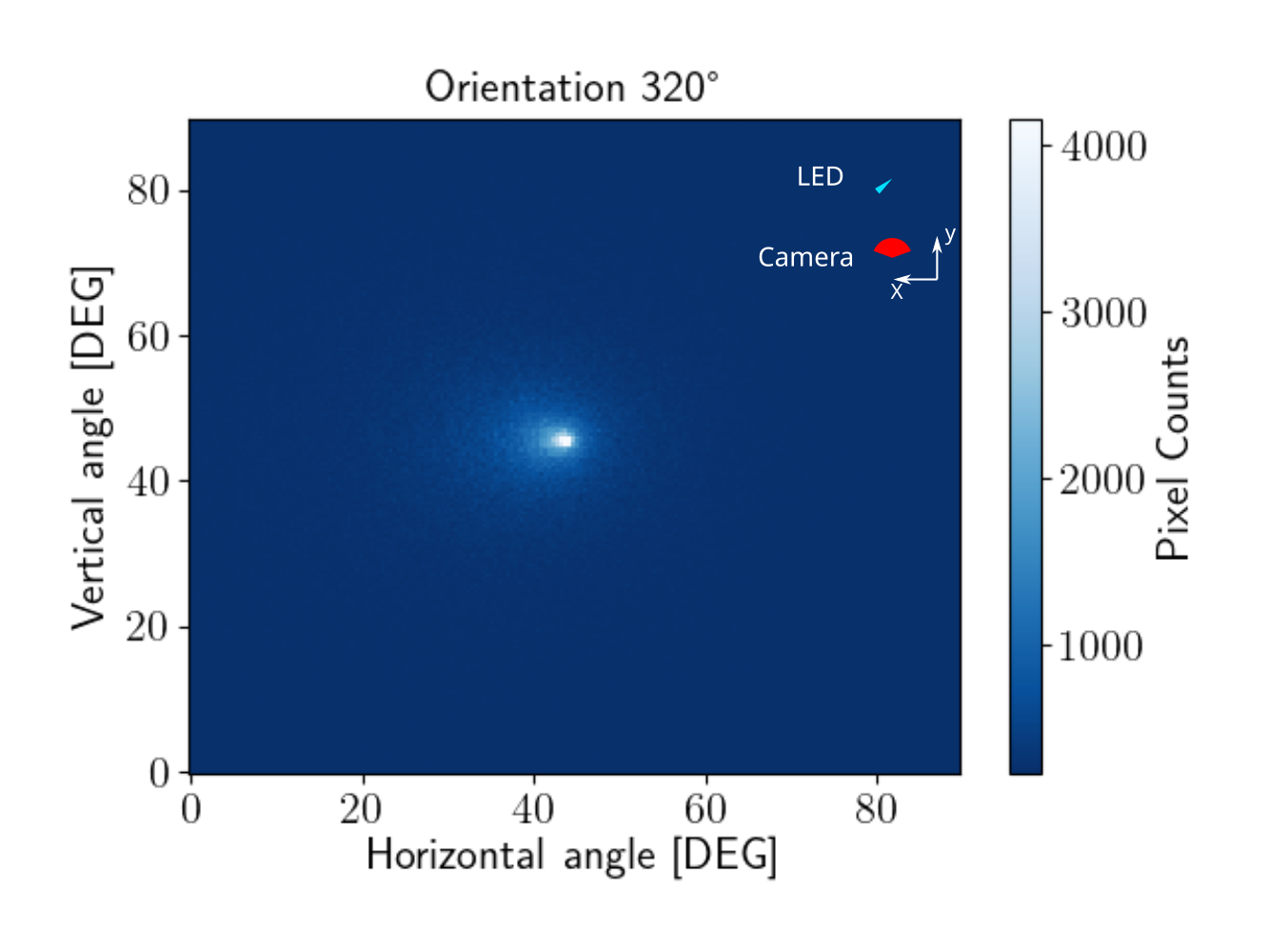}
        \caption[margin=0.8cm]{Simulated camera images at different LED orientations. The LED is directly facing the camera at 270 degrees and has a profile width of 22.5 degrees. Noise has been added to these images according to the camera noise model in Figure \ref{fig:camera_noise}.}
    \label{fig:LED_orientation}
\end{figure}

In Figure~\ref{fig:LED_orientation} simulated images for different LED orientation can be seen. The images show the LED pointing in different directions as the LED is turned in the simulation. These images are relevant for bulk ice property analyses to separately estimate the LED orientation to minimize the effect of the orientation as a systematic uncertainty on the bulk ice measurements.

\section{Conclusion}\label{sec5}

A new framework has been developed to simulate images of the camera systems in the IceCube Upgrade and other camera devices. The framework simplifies the production of simulated images and illustrates the visible effects of the ice properties on the expected light signatures for camera based calibration systems in ice and water.


\bibliographystyle{ICRC}
\bibliography{ICRC2023_CamSim}
\clearpage
\input{authorlist_IceCube.tex}

\end{document}

%% file: authorlist_IceCube.tex
 \section*{Full Author List: IceCube Collaboration}

\scriptsize
\noindent
R. Abbasi$^{17}$,
M. Ackermann$^{63}$,
J. Adams$^{18}$,
S. K. Agarwalla$^{40,\: 64}$,
J. A. Aguilar$^{12}$,
M. Ahlers$^{22}$,
J.M. Alameddine$^{23}$,
N. M. Amin$^{44}$,
K. Andeen$^{42}$,
G. Anton$^{26}$
% ...
\\
\\
% ...
$^{12}$ Universit{\'e} Libre de Bruxelles, Science Faculty CP230, B-1050 Brussels, Belgium \\
% ...
$^{17}$ Department of Physics, Loyola University Chicago, Chicago, IL 60660, USA \\
$^{18}$ Dept. of Physics and Astronomy, University of Canterbury, Private Bag 4800, Christchurch, New Zealand \\
% ...
$^{22}$ Niels Bohr Institute, University of Copenhagen, DK-2100 Copenhagen, Denmark \\
$^{23}$ Dept. of Physics, TU Dortmund University, D-44221 Dortmund, Germany \\
% ...
$^{26}$ Erlangen Centre for Astroparticle Physics, Friedrich-Alexander-Universit{\"a}t Erlangen-N{\"u}rnberg, D-91058 Erlangen, Germany \\
% ...
$^{40}$ Dept. of Physics and Wisconsin IceCube Particle Astrophysics Center, University of Wisconsin{\textendash}Madison, Madison, WI 53706, USA \\
% ...
$^{42}$ Department of Physics, Marquette University, Milwaukee, WI, 53201, USA \\
% ...
$^{44}$ Bartol Research Institute and Dept. of Physics and Astronomy, University of Delaware, Newark, DE 19716, USA \\
% ...
$^{63}$ Deutsches Elektronen-Synchrotron DESY, Platanenallee 6, 15738 Zeuthen, Germany  \\
$^{64}$ Institute of Physics, Sachivalaya Marg, Sainik School Post, Bhubaneswar 751005, India \\
% ...

\subsection*{Acknowledgements}

\noindent
The authors gratefully acknowledge the support from the following agencies and institutions:
USA {\textendash} U.S. National Science Foundation-Office of Polar Programs,
U.S. National Science Foundation-Physics Division,
U.S. National Science Foundation-EPSCoR,
Wisconsin Alumni Research Foundation,
Center for High Throughput Computing (CHTC) at the University of Wisconsin{\textendash}Madison,
Open Science Grid (OSG),
Advanced Cyberinfrastructure Coordination Ecosystem: Services {\&} Support (ACCESS),
Frontera computing project at the Texas Advanced Computing Center,
U.S. Department of Energy-National Energy Research Scientific Computing Center,
Particle astrophysics research computing center at the University of Maryland,
Institute for Cyber-Enabled Research at Michigan State University,
and Astroparticle physics computational facility at Marquette University;
Belgium {\textendash} Funds for Scientific Research (FRS-FNRS and FWO),
FWO Odysseus and Big Science programmes,
and Belgian Federal Science Policy Office (Belspo);
Germany {\textendash} Bundesministerium f{\"u}r Bildung und Forschung (BMBF),
Deutsche Forschungsgemeinschaft (DFG),
Helmholtz Alliance for Astroparticle Physics (HAP),
Initiative and Networking Fund of the Helmholtz Association,
Deutsches Elektronen Synchrotron (DESY),
and High Performance Computing cluster of the RWTH Aachen;
Sweden {\textendash} Swedish Research Council,
Swedish Polar Research Secretariat,
Swedish National Infrastructure for Computing (SNIC),
and Knut and Alice Wallenberg Foundation;
European Union {\textendash} EGI Advanced Computing for research;
Australia {\textendash} Australian Research Council;
Canada {\textendash} Natural Sciences and Engineering Research Council of Canada,
Calcul Qu{\'e}bec, Compute Ontario, Canada Foundation for Innovation, WestGrid, and Compute Canada;
Denmark {\textendash} Villum Fonden, Carlsberg Foundation, and European Commission;
New Zealand {\textendash} Marsden Fund;
Japan {\textendash} Japan Society for Promotion of Science (JSPS)
and Institute for Global Prominent Research (IGPR) of Chiba University;
Korea {\textendash} National Research Foundation of Korea (NRF);
Switzerland {\textendash} Swiss National Science Foundation (SNSF);
United Kingdom {\textendash} Department of Physics, University of Oxford.

%% file: ICRC2023_CamSim.bbl
\providecommand{\href}[2]{#2}\begingroup\raggedright\begin{thebibliography}{10}

\bibitem{ICRC2019:ICU-project}
{\bfseries IceCube} Collaboration, A.~Ishihara
  \href{http://dx.doi.org/10.22323/1.358.1031}{{\em PoS} {\bfseries ICRC2019}
  (2021) 1031}.

\bibitem{ICdetector}
{\bfseries IceCube} Collaboration, { M. G. Aartsen et. al.}
  \href{http://dx.doi.org/10.1088/1748-0221/12/03/P03012}{{\em JINST}
  {\bfseries 12} no.~03, (2017) P03012}.

\bibitem{ICRC2021:D-Egg}
{\bfseries IceCube} Collaboration, C.~Hill {\em PoS} {\bfseries ICRC2021}
  (2021) 1042.

\bibitem{ICRC2021:mDOM}
{\bfseries IceCube} Collaboration, L.~Classen {\em PoS} {\bfseries ICRC2021}
  (2021) 1070.

\bibitem{ICRC2019:anisotropy}
{\bfseries IceCube} Collaboration, D.~Chirkin and M.~Rongen
  \href{http://dx.doi.org/10.22323/1.358.0854}{{\em PoS} {\bfseries ICRC2019}
  (2020) 854}.

\bibitem{SPICE}
{\bfseries IceCube} Collaboration, M.~G. Aartsen {\em et~al.}
  \href{http://dx.doi.org/10.1016/j.nima.2013.01.054}{{\em Nucl. Instrum. Meth.
  A} {\bfseries 711} (2013) 73--89}.

\bibitem{ICRC2019:ICUcamera}
{\bfseries IceCube} Collaboration, W.~Kang, C.~T\"onnis, and C.~Rott
  \href{http://dx.doi.org/10.22323/1.358.0928}{{\em PoS} {\bfseries ICRC2019}
  (2020) 928}.

\bibitem{ICRC2017:Gen2camera}
{\bfseries IceCube-Gen2} Collaboration, M.~Jeong and W.~Kang
  \href{http://dx.doi.org/10.22323/1.301.1040}{{\em PoS} {\bfseries ICRC2017}
  (2018) 1040}.

\bibitem{ppc}
{\bfseries IceCube} Collaboration, {D. Chirkin}
  \href{http://dx.doi.org/10.1016/j.nima.2012.11.170}{{\em Nucl. Instrum. Meth.
  A} {\bfseries 725} (2013) 141--143}.

\bibitem{Clsim:2022yy}
{H. Schwanekamp, R. Hohl, D. Chirkin et al.}
  \href{http://dx.doi.org/https://doi.org/10.1007/s41781-022-00080-8}{{\em
  Comput Softw Big Sci} {\bfseries 6} (2022) 4}.

\bibitem{Aartsenosc}
{\bfseries IceCube} Collaboration, M.~G.~A. et. al.
  \href{http://dx.doi.org/10.1103/physrevlett.120.071801}{{\em Physical Review
  Letters} {\bfseries 120} no.~7, (Feb, 2018) }.

\bibitem{refId0}
{\bfseries IceCube} Collaboration, M.~Rongen
  \href{http://dx.doi.org/10.1051/epjconf/201611606011}{{\em EPJ Web of
  Conferences} {\bfseries 116} (2016) 06011}.

\end{thebibliography}\endgroup
